\documentclass[doublespace,onecolumn]{SSA-SRL}

\usepackage{booktabs}
\usepackage[authormarkup=none]{changes}
\definechangesauthor[name={hn}, color=blue]{HN}
\definechangesauthor[name={bb}, color=red]{BB}
\definechangesauthor[name={r1}, color=red]{R1}

\citethisauthor{J.~Liu, H.~Li, and S.~Yuan}
\doi{00.0000/000000000}
\recdate{00 Month 0000}

\begin{document}

\title{Characterizing Vehicle-Induced Distributed Acoustic Sensing Signals for Accurate Urban Near-Surface Imaging}

\author[*1,2\orc{0000-0001-5559-1627}]{Jingxiao Liu}
\author[1\orc{0000-0002-3230-8339}]{Haipeng Li}
\author[1\orc{0000-0001-8922-9423}]{Siyuan Yuan}
\author[2\orc{0000-0002-7998-3657}]{Hae Young Noh}
\author[1\orc{0000-0002-0961-2287}]{Biondo Biondi}

\affil[1]{Department of Geophysics, Stanford University, Stanford, California, United States}{~Mailing address: 397 Panama Mall Mitchell Building, 3rd Floor, Stanford, CA 94305, United States}
\affil[2]{Department of Civil and Environmental Engineering, Stanford, California, United States}{~Mailing address: Y2E2, 473 Via Ortega, Room 311, Stanford, CA 94305, United States}
\corau{*Corresponding author: jingxiao@mit.edu}

\maketitle

\section{Abstract}
\par
Continuous seismic monitoring of the near-surface structure is crucial for urban infrastructure safety, aiding in the detection of sinkholes, subsidence, and other seismic hazards. 
Utilizing existing telecommunication optical fibers as Distributed Acoustic Sensing (DAS) systems offers a cost-effective method for creating dense seismic arrays in urban areas.
DAS leverages roadside fiber-optic cables to record vehicle-induced surface waves for near-surface imaging.
However, the influence of roadway vehicle characteristics, such as vehicle weight/size and driving speed, on their induced surface waves and the resulting imaging of near-surface structures is poorly understood. 
In this study, we investigate surface waves generated by vehicles of varying weights and speeds to provide insights into accurate and efficient near-surface characterization.
We first classify vehicles into light, mid-weight, and heavy based on the maximum amplitudes of quasi-static DAS records. 
Vehicles are also classified by their traveling speed (slow, medium, and fast) using their arrival times at DAS channels. 
To investigate how vehicle characteristics influence the induced surface waves, we extract phase velocity dispersion and invert the subsurface structure for each vehicle class by retrieving virtual shot gathers (VSGs).
Our results reveal that heavy vehicles produce higher signal-to-noise ratio surface waves, and a sevenfold increase in vehicle weight can reduce uncertainties in phase velocity measurements from dispersion spectra by up to 3$\times$, particularly at lower frequencies.
Thus, data from heavy vehicles better constrain structures at greater depths. 
Additionally, with driving speeds ranging from 5 to 30 meters per second in our study, differences in the dispersion curves due to vehicle speed are less pronounced than those due to vehicle weight. 
Our results suggest judiciously selecting and processing surface wave signals from certain vehicle types can improve the quality of near-surface imaging in urban environments.

\section{Introduction}
\par
Continuous seismic monitoring of the near-surface structure is essential for ensuring the safety and stability of urban infrastructure~\citep{https://doi.org/10.1046/j.1365-2478.1999.00138.x,https://doi.org/10.1029/2019GL086115}. It plays a crucial role in the early detection of potential hazards such as sinkholes, subsidence, and other ground instabilities that can pose significant risks to buildings, roads, and utilities.
While conventional methods, such as controlled source surveys~\citep{doi:10.1061/(ASCE)1090-0241(1998)124:8(757),doi:10.1190/1.1444590} and ambient noise interferometry~\citep{doi:10.1126/science.1108339, doi:10.1190/1.3457445}, can provide near-surface images, repeatedly conducting such surveys or deploying dense seismic arrays at urban scales is challenging regarding logistics and operational expenses. 

% While conventional methods, such as controlled source surveys~\citep{doi:10.1061/(ASCE)1090-0241(1998)124:8(757),doi:10.1190/1.1444590} can provide high-quality near-surface images, repeatedly conducting such surveys is challenging regarding logistics and operational expenses. Ambient noise interferometry in a cost-effective way for subsurface characterization ~\citep{doi:10.1126/science.1108339, doi:10.1190/1.3457445}, while its quality relies on the dense seismic arrays and even distribution of the sources in the urban. 

\par
Distributed Acoustic Sensing (DAS) has emerged as a promising technology that repurposes telecommunication optical fibers as dense seismic arrays with low-cost and low-maintenance~\citep{10.1190/tle36121025.1,10.1785/0220190112,10.1190/tle39090646.1,annurev:/content/journals/10.1146/annurev-earth-072420-065213}.
DAS-based monitoring is achieved by connecting an interrogator to one end of a standard telecommunication optical fiber. 
The interrogator injects short laser pulses into the fiber and detects the phase shifts of Rayleigh scattered lights that are linearly related to the axial strain along the fiber and return at the predicted travel time~\citep{10.1063/1.4939482}. 
This technique allows the measurement of the strain field caused by surrounding disturbance (e.g., earthquakes, moving vehicles) on the fiber with meter-scale spatial resolution over tens of kilometers. 
Integrated with ambient noise interferometry, DAS data collected from existing and extensive ``dark fiber" telecommunication infrastructure has been used to efficiently and cost-effectively image the near-surface structure~\citep{2437101751,Ajo-Franklin2019}.
\par
In urban areas, recent studies have utilized {ambient noise interferometry to analyze} DAS data collected from ``dark fiber" to characterize near-surface properties such as S-wave velocity and attenuation~\citep{Dou2017,10.1190/tle39090646.1,https://doi.org/10.1029/2021GL096503,10.1785/0220240050,10.1785/0220230104}. 
This approach is feasible because most telecommunication optical fibers in urban areas are deployed along roadsides, making traffic the primary seismic source for interferometry. 
{However, conventional ambient noise interferometry has limitations, including long cross-correlation periods required for convergence, diminished temporal resolution, and challenges in capturing high-frequency signals, which are often attenuated in ambient noise~\citep{https://doi.org/10.1029/2023JB028033}. These issues are exacerbated by the heterogeneous and localized nature of noise sources in urban areas, which contrasts with the assumption of uniformly distributed ambient sources typically required for ambient noise interferometry.} 
In particular, roadway vehicles have distinct characteristics, such as vehicle weight, size, and driving speed, that impact the induced surface waves and the inversion of near-surface structures.
Directly analyzing surface wave signals from vehicles of varying weights and speeds may result in low-quality near-surface imaging. 
{For instance, we observed that picked fundamental mode dispersion curves from vehicle-induced DAS signals are robust at frequencies above 5 Hz but not at lower frequencies. This is because the dominant traffic type, small/light vehicle, primarily excites surface waves at frequencies above 5 Hz.}
Therefore, understanding the effects of vehicle characteristics on the induced surface waves is important for improving the quality of near-surface imaging using vehicle-induced DAS signals in urban environments.

\par
In this study, we characterize surface waves generated by moving vehicles of different weights/sizes (sizes are closely correlated with weights) and speeds in the Stanford DAS-2 experiment.
Building on our recent work~\citep{https://doi.org/10.1029/2023JB028033}, we first use a specialized Kalman filter algorithm~\citep{10.1145/3596262} to track vehicle locations and then classify vehicles by the peak values of their average quasi-static signals into weight and speed categories
{: light-weight (smaller than the mode value; peak value $\le$0.57), heavy-weight (laying on the distribution tail; peak value $>$1.2), and mid-weight (in between; 0.57 $<$ peak value $\le$1.2),.
Vehicles are also categorized by their arrival times at DAS into traveling speed categories -- slow, mid-speed, and fast -- using the mean speed plus or minus one standard deviation as thresholds. }
Vehicle-induced surface wave windows are then selected using the tracked vehicle trajectories.
Virtual shot gathers (VSGs) for the same class of vehicles are retrieved from their surface wave windows and stacked to increase the signal-to-noise ratio~\citep{https://doi.org/10.1029/2023JB028033}.
A VSG is a shot gathered as if a seismic source were excited at a specific location. 
This allows us to analyze the phase velocity dispersion at the location and further provides insights into how vehicle characteristics influence the induced surface waves.
We also quantify the uncertainties in the picked dispersion curves as the number of stacked VSGs increases by bootstrapping the standard deviation of the phase velocities. 
{Our results show that stacking VSGs reduces the uncertainty of picked dispersion curves, with convergence occurring after signals from approximately 55 vehicles are stacked. }
Heavy-weight vehicles show higher signal-to-noise ratios and reduced uncertainty in phase velocity estimates, particularly at lower frequencies (e.g., <5 Hz), allowing for better constraints of near-surface structures at greater depths. 
We also find that differences in the picked dispersion curves for vehicles with different speeds are less pronounced than those due to vehicle weight.
{In particular, for fine-grained temporal monitoring (e.g., daily monitoring) of near-surface characteristics, the low-frequency phase velocities (e.g., $<$5 Hz) should be derived from stacked heavy-weight vehicle VSGs and the high-frequency velocities from stacked VSGs of all classes of vehicles until convergence is achieved; for coarse-grained monitoring (e.g., weekly monitoring), the best dispersion curve should come from stacked heavy-weight vehicle VSGs, which is more efficient and accurate.}
\par
{In summary, our results suggest that judiciously selecting certain classes of vehicles can be beneficial in processing vehicle-induced surface waves. The proposed approach in this study offers three key advantages:}
\begin{enumerate}
    \item {By stacking the results of automatically tracked vehicles, we achieve highly accurate VSGs that capture detailed subsurface information with a high temporal resolution of up to daily intervals, a level of resolution challenging to obtain with conventional ambient noise interferometry.}
    \item {Targeting specific vehicle classes further improves the signal-to-noise ratio in VSGs and enhances dispersion analysis accuracy, especially at lower frequencies. This leads to improved phase velocity estimation and greater sensitivity for imaging deeper subsurface structures.}
    \item {The computational cost is minimized by only cross-correlation to the surface wave windows generated by the selected vehicle classes, significantly accelerating the processing workflow for large-scale applications.}
\end{enumerate}

\section{Methods}
\par
In this section, we introduce the methodology employed to investigate the influence of vehicle characteristics on the induced surface waves. 
We begin by tracking and characterizing vehicles from roadside DAS recordings to classify them based on attributes, including weight/size and traveling speed. 
These tracking results also enable us to select vehicle-induced surface wave windows. 
We then introduce a targeted interferometry workflow to retrieve VSGs from these vehicle-induced surface wave windows, enabling us to analyze the phase velocity dispersion across different vehicle types. 
Finally, we present the inversion method used for estimating subsurface S-wave velocity, which allows us to assess how vehicle characteristics affect the results of inverted structures.

\subsection{Surface wave window selection based on vehicle tracking and characterization}

\par
As vehicles travel along a roadside DAS array, they produce two types of signals: quasi-static deformation signals (usually below 1 Hz and influenced by vehicle velocity) resulting from their weight and vehicle-induced surface waves (1-30 Hz), primarily Rayleigh waves ~\citep{10.1190/segam2021-3584136.1}, resulting from the vehicle-road interactive dynamics. 
In our study, the quasi-static deformation signals are utilized to track the locations of moving vehicles on the DAS recordings using a specialized Kalman filter and smoothing algorithm~\citep{10.1145/3596262}.
\par
The algorithm estimates the vehicles' arrival times at each DAS channel through a prominence-based peak detection method. 
Peak prominence measures the extent to which a peak stands out due to its height and relative location to other peaks. 
Vehicle position tracking is then formulated as a spatial-domain state-space model, considering the arrival time of a vehicle at every DAS channel as a measurement. 
The vehicle arrival time at the $k$-th DAS channel, $t$, and its derivative, $\dot{t}$, are represented by the linear state space ${\bf t}_k = [t,\dot{t}]^T$. 
Here, $\dot{t}$ denotes the slowness. 
According to the physical law of motion, the state-space model is given by
\begin{equation}
\begin{aligned}
    & {\bf t}_k={\bf A}_k{\bf t}_{k-1}+{\bf w}_k\\
    & {z}_k={\bf C}{\bf t}_k+{ v}_k,
\end{aligned}
\end{equation}
where
$${\bf A}_k = \begin{bmatrix}1 & \Delta x_k\\
0 & 1\end{bmatrix},~{\bf C}=[1,~0],~{\bf w}_k\sim\mathcal{N}(0,{\bf Q}_k),~{ v}_k\sim\mathcal{N}(0,\sigma_z),$$
% $${\bf Q}_k = \begin{bmatrix}\frac{1}{4}\Delta x_k^4 & \frac{1}{2}\Delta x_k^3\\
% \frac{1}{2}\Delta x_k^3 & \Delta x_k^2\end{bmatrix}\sigma_{\ddot{t}}^2,$$
and $\Delta x_k$ is the distance between the $(k-1)$-th and the $k$-th DAS channel projected on the road; ${z}_k$ is the observed vehicle arrival time at DAS channel $k$; ${\bf w}_k$ is the process noise; ${\bf v}_k$ is the observation noise; and ${\bf Q}_k$ and $\sigma_z$ are the standard deviations of the process and the observation noise, respectively. 
The Kalman filter and smoothing algorithm are then developed to recursively determine the posterior probability of vehicle arrival times across space (in the direction of vehicle motion) by integrating spatial-dependent vehicle detection results from multiple DAS channels. 
This algorithm includes a forward filtering pass and a backward smoothing pass. 
The forward pass estimates the posterior probability of vehicle arrival time over space using detection results from previous and current DAS channels. 
% Given the absence of direct measurements (i.e., vehicle arrival time) at each DAS channel, the algorithm includes a measurement prediction phase that predicts the measurement, ${z}_k$, by identifying the detection in the current DAS channel with the highest probability of the predicted arrival time:
% \begin{equation}
% \label{eq:bayes}
%     \hat{z}_k=\arg\max_{z\in {\bf D}_k} f(z;\hat{t}_{k|k-1}, \sigma_{t,k|k-1})
% \end{equation}
% where ${\bf D}_k$ are the peak detection results for DAS channel $k$; $f(z)$ is the probability density function of the predicted arrival time following a Gaussian distribution: $\mathcal{N}(z;\hat{{t}}_{k|k-1},\sigma_{t,k|k-1}^2)$; $\hat{t}_{k|k-1}$ and $\sigma_{t,k|k-1}$ are the predicted arrival time estimate and its standard deviation, respectively. 
The backward pass employs the Rauch-Tung-Striebel smoother~\citep{doi:10.2514/3.3166} to estimate the posterior probability of vehicle arrival time using detection results from all DAS channels. 
Vehicle motion state estimation or tracking is ultimately achieved by converting the arrival time and its derivative at every DAS channel, $[t,\dot{t}]$, into vehicle position and speed, $[x,v]$, using the geographic locations of the DAS channels.
\par
Additionally, vehicle weights are estimated using the quasi-static component of the DAS signals. 
Assuming linear elasticity of the road and near-surface structures, the prominence amplitude of the quasi-static signal is approximately proportional to the vehicle's weight. 
Heavier vehicles produce larger prominence amplitudes. 
Therefore, vehicles' relative weights are estimated by computing the average of the quasi-static signal prominence, $\frac{1}{K}\sum_{k=1}^K P_k$, where $P_k$ is the prominence amplitude {(or peak value)} of the vehicle's quasi-static signal at DAS channel $k$. 
The vehicle's axle length and configuration can also be estimated from its quasi-static signal~\citep{10.1145/3596262,10280620}. 
Yet, since vehicle length and weight are closely correlated, this study simplifies the analysis without considering different axle lengths and configurations.
\par
The tracked vehicle trajectories on the DAS signals allow for the selection of corresponding surface wave windows induced by vehicle movement. 
To avoid interference from nearby vehicles, only isolated vehicles with a minimum separation of 25 seconds from other vehicles are selected. 
For each isolated vehicle, a 25-second and 500-meter window centered on the vehicle's trajectory is chosen. 
The DAS signal at the center of each surface wave window is considered the pivot trace for constructing virtual shot gathers, as described in the following section.
Based on the vehicle tracking and characterization results, the selected vehicle-induced surface wave windows are classified into different classes according to vehicle weight (which correlates with vehicle size) and traveling speed. 
This classification helps to study the effects of vehicle characteristics on the generated surface waves and their dispersion curves. 

\subsection{Retrieval of virtual shot gathers by targeted interferometry}
\par
To characterize seismic wave propagation, we retrieve VSGs from the selected vehicle-induced surface waves by cross-correlating the pivot trace with other traces within the selected surface wave windows~\citep{https://doi.org/10.1029/2023JB028033}. 
The signals in the vehicle-induced surface wave windows contain both forward- and backward-propagating waves on either side of the vehicle's trajectory. 
We handle the positive and negative offset sections of wave interferometry separately.
\par
For the backward-propagating waves, i.e., those traveling in the direction opposite to the vehicle movement, we define the cross-correlation function $C\left(\mathbf{x}_s, \mathbf{x}_r, \tau \right)$ as follows:
\begin{equation}
C\left(\mathbf{x}_s, \mathbf{x}_r, \tau \right) = \left\{\begin{array}{l}
\int_{t_s+\epsilon}^{t_s+\epsilon+\Delta t} u\left(t+\tau; \mathbf{x}_r\right) \cdot u\left(t; \mathbf{x}_s\right) d t, \quad \mathbf{x}_r < \mathbf{x}_s \\
\\
\int_{t_r+\epsilon}^{t_r+\epsilon+\Delta t} u\left(t-\tau; \mathbf{x}_r\right) \cdot u\left(t; \mathbf{x}_s\right) d t, \quad \mathbf{x}_r \geq \mathbf{x}_s,
\end{array}\right.
\end{equation}
where $u\left(t; \mathbf{x}\right)$ represents the recorded DAS trace at time $t$ and location $\mathbf{x}$, and $C\left(\mathbf{x}_s, \mathbf{x}_r, \tau \right)$ denotes the inter-channel correlation between recorded DAS strain wavefields at two channel pairs $\mathbf{x}_s$ and $\mathbf{x}_r$ with $\tau$ being the time lag. 
Here, $C\left(\mathbf{x}_s, \mathbf{x}_r, \tau \right)$ approximates the wavefield as if a source is placed at $\mathbf{x}_s$ and a receiver at $\mathbf{x}_r$.
For the negative-offset section in VSGs, channels traversed by the vehicle ($\mathbf{x}_r < \mathbf{x}_s$) are cross-correlated with the pivot trace $u(t, \mathbf{x}_s)$ within the time window $[t_s+\epsilon, t_s+\epsilon+\Delta t]$. 
Here, $ t_s $ denotes the vehicle's arrival time at the virtual source location $ \mathbf{x}_s $, $\Delta t$ represents the selected time window length for cross-correlation, and $\epsilon$ is a time lag introduced to avoid near-field effects. 
For the positive-offset section of VSGs, the cross-correlation is performed in the time window $[t_r+\epsilon, t_r+\epsilon+\Delta t]$, where $t_r$ represents the arrival time of the traveling vehicle at the virtual receiver location $\mathbf{x}_r$.
For forward-propagating waves, the time window $[t_r-\epsilon-\Delta t, t_r-\epsilon]$ is used for the negative-offset section ($\mathbf{x}_r < \mathbf{x}_s$), and $[t_s-\epsilon-\Delta t, t_s-\epsilon]$ for the positive-offset segment ($\mathbf{x}_r \geq \mathbf{x}_s$).
\par
We stack (or sum) VSGs from vehicles within the same class (based on weight/size and traveling speed) at the same location. 
This stacking process improves the signal-to-noise ratio by reducing random noise. 
Subsequently, we apply the phase-shift method~\citep{doi:10.1190/1.1820161} to measure dispersion for VSGs stacked from vehicles with various characteristics.

\subsection{Inversion approach}
\par
We use the software ``evodcinv"~\citep{luu_2023_8007868} to invert for near-surface S-wave velocity based on classified dispersion measurements.
The root mean square error is used as the misfit function $\mathcal{X}({\bf m})$ for a test layered model ${\bf m}$:
\begin{equation}
\mathcal{X}({\bf m})=\sqrt{\frac{1}{N}\sum_i^N\Big[\frac{c_i({\bf m})-c_i^{obs}}{\sigma_i}\Big]^2},
\end{equation}
where $c_i({\bf m})$ and $c_i^{obs}$ denote the synthetic and observed phase/group velocities at specific frequencies and modes, respectively, $\sigma_i$ is the corresponding standard deviation, and $N$ is the total number of layers.
\par
To solve this non-linear optimization problem, we utilize the Competitive Particle Swarm Optimization (CPSO) method~\citep{2890425708,luu_2023_8007868}. 
CPSO enhances the basic Particle Swarm Optimization (PSO) algorithm by incorporating a ‘competition’ mechanism, which increases swarm diversity and helps escape local minima. 
Detailed information on the CPSO solver can be found in~\cite{luu_2023_8007868}. 
Following empirical recommendations ~\citep{870279}, we set inertia weight parameter $w$ and two acceleration parameters $\phi_p$ and $\phi_g$ as follows: $w = 0.7298$ and $\phi_p = \phi_g = 1.49618$. 

\section{Urban DAS Data and Preprocessing}

\begin{figure}[!ht]
    \centering
    \includegraphics[width=1\linewidth]{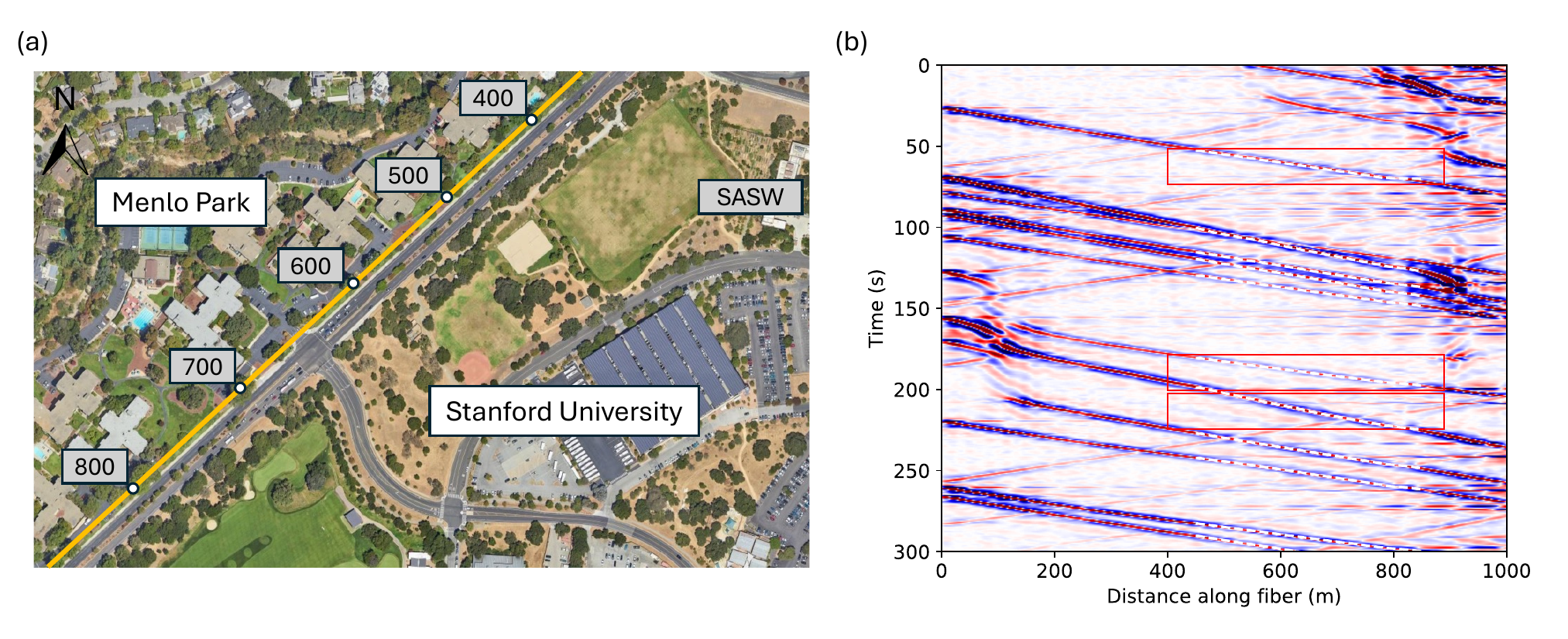}
    \caption{The DAS array section used in this study from the Stanford-2 DAS experiment and the DAS recording after preprocessing.
    (a) Map view of a roadside section of the Stanford DAS-2 Array. 
    Distances along the fiber are labeled on the map. 
    Menlo Park is located to the west of the DAS array, and Stanford University is located to the east. 
    The Spectral Analysis of Surface Waves (SASW) survey, performed in 2009, is approximately 200 meters east of our DAS array.
    (b) A five-minute section of the DAS recordings. 
    The dashed white lines indicate vehicle trajectories estimated using our Kalman filter and smoothing algorithm. 
    The red boxes highlight the selected surface wave windows with isolated vehicles.}
    \label{fig:map}
\end{figure}

% This section describes the DAS array and the recorded DAS signals used in this study, followed by an analysis of the data to quantify the uncertainty of the estimated dispersion curves. 
% The uncertainty quantification compares and characterizes the surface waves generated by vehicles with different properties.
% \subsection{DAS array and recordings}
\par
The data for this study was collected during the Stanford DAS-2 experiment, conducted at Stanford University, which has been continuously recording data since December 10th, 2019. 
Data acquisition was performed using an ODH-3 interrogator from Luna--OptaSense, operating at a sampling rate of 250 samples per second with a gauge length of 16 meters. 
The DAS array comprises 1250 channels, each spaced 8.16 meters apart. 
We utilized a section of the fiber along Sand Hill Road, as illustrated in Figure~\ref{fig:map}a. 
This section was selected due to its complex and varied near-surface conditions. 
The DAS channels were geo-located through controlled driving experiments~\citep{10.1190/segam2021-3584136.1}. 
This study focuses on DAS data collected over a half-month period, from December 2nd to December 17th, under similar weather and climate conditions.
We focus on data collected between midnight and 6 am, as the lower traffic during these hours provides more isolated vehicle signals.
Additionally, a Spectral Analysis of Surface Waves (SASW) survey~\citep{Wong_Stokoe_2009} performed in 2009 is located approximately 200 meters east of our DAS array, which is used to justify our inversion results.
\par
The DAS signals are processed differently for vehicle tracking and near-surface characterization. 
Initially, a band-pass filter (0.1--1 Hz) was applied to obtain the quasi-static signals for vehicle tracking. 
The selected vehicle-induced DAS signals are band-pass filtered between 1 and 30 Hz to isolate the surface wave windows.  
This frequency band is selected because the gauge length influences the high-frequency dispersion spectra, acting as a spatial low-pass filter~\citep{yuan2023large}. 
The 16 m gauge length used in this study restricts the recording of high-frequency signals. 
Shorter gauge lengths would enable the capture of energy at higher frequencies in the signal.
Figure~\ref{fig:map}b presents an example of the DAS signals, with the selected surface wave windows marked by red boxes. These selected windows are then used to construct VSGs.

\section{Results}
\par
This section presents the vehicle-induced DAS signals from vehicles of different characteristics, followed by the corresponding dispersion analysis. 
We compare the picked dispersion curves obtained from surface wave windows induced by vehicles with various weights/sizes and traveling speeds. 
Additionally, we show the inversion of shear-wave velocity using the estimated dispersion curves and further validate our results with a previous SASW survey.

\subsection{Characterization of vehicle-induced DAS signal }
\par
\begin{figure}
    \centering
    \includegraphics[width=1\linewidth]{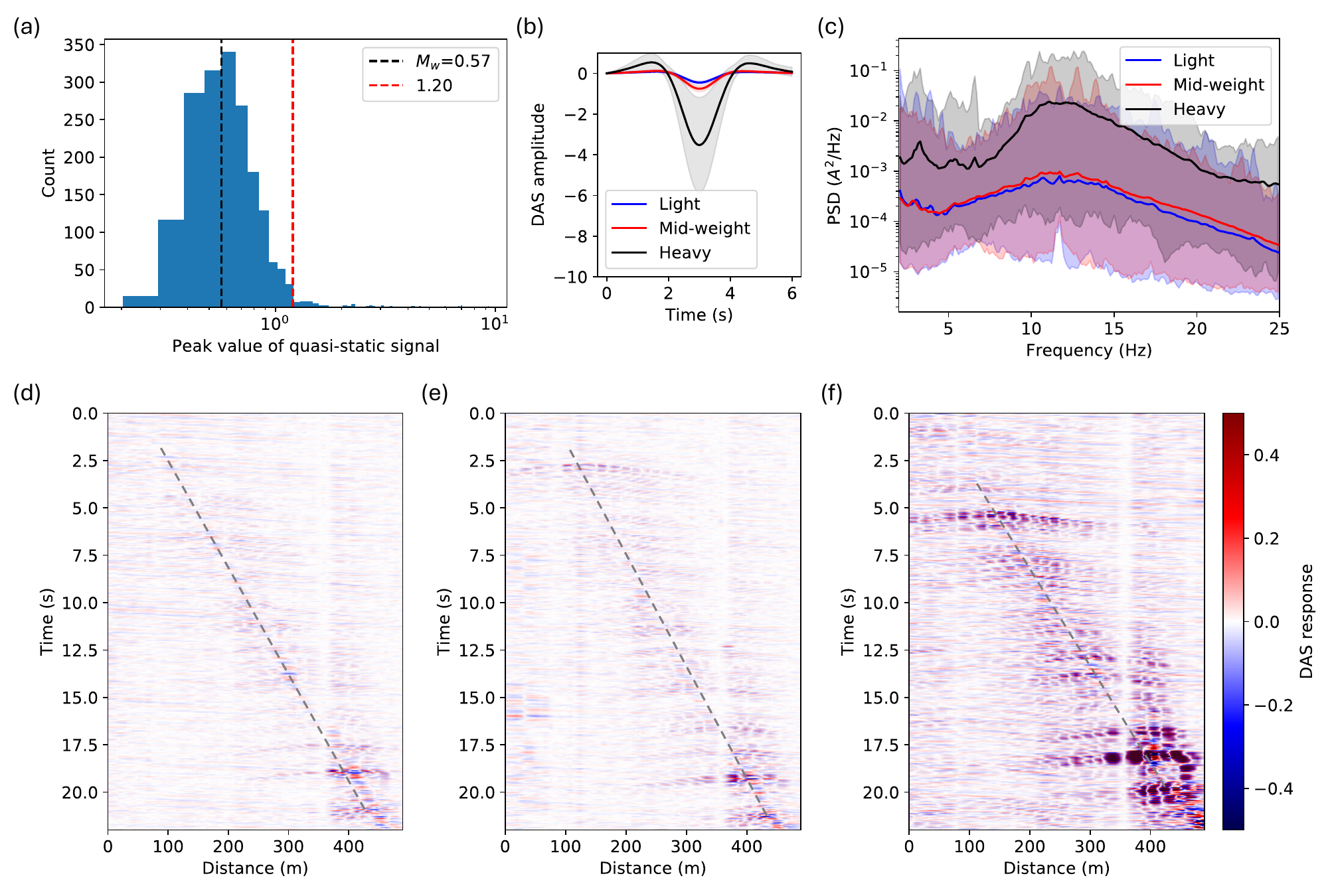}
    \caption{DAS signals induced by vehicles of different weights. 
    (a) The distribution of the peak values of the averaged quasi-static signals (on a logarithmic scale). 
    The peak values are used to represent the weights of vehicles.
    Thresholds at the mode (0.57) and at the tail (1.2) of peak values differentiate between light, mid-weight, and heavy vehicles. 
    (b) The quasi-static DAS signals induced by classified vehicles of different weights. 
    The solid curves represent the mean signals, and the shaded areas indicate the signal ranges. 
    (c) The power spectral density (PSD) of surface wave signals in the frequency band of 1--25 Hz for vehicles of different weights. 
    The solid curves show the mean PSDs, and the shaded areas indicate the PSD ranges. 
    (d)-(f) Examples of surface wave signals (1--30 Hz) for vehicles of different weights plotted with the same color map. 
    The color version of this figure is available only in the electronic edition.}
    \label{fig:size}
\end{figure}

We begin by characterizing the vehicle-induced DAS signals based on the estimated weights/sizes of the vehicles. 
To control for the variable of traveling speed, we select vehicles traveling within one standard deviation of the mode speed of all vehicles. 
We extract the quasi-static signals for each DAS channel by isolating a time window from the band-passed DAS signal (0.1–1 Hz), centered on the vehicle's arrival time at that channel.
Figure~\ref{fig:size}a displays the histogram of peak values of the average quasi-static signals of these vehicles.
{The peak value is defined as the average prominence amplitude of the vehicle’s quasi-static signal across all DAS channels within each surface wave window.}
We categorize the vehicles into three classes based on the peak values of their average quasi-static signals: vehicles with peak values smaller than the mode (which equals 0.57) are classified as light; those with peak values laying on the distribution tail (peak value $>1.2$) are classified as heavy; and those with values in between these thresholds are considered mid-weight. 
During the half-month period, we recorded 734 light, 1,058 mid-weight, and 103 heavy vehicles, respectively. 
Heavy vehicles, on average, weigh approximately seven times more than light vehicles.
Figure~\ref{fig:size}b illustrates the average quasi-static DAS signals for vehicles having different weights, where the shaded areas highlight the signal ranges.
Figure~\ref{fig:size}c shows the power spectral densities (PSD) of the surface wave signals band-passed between 1 and 30 Hz for different vehicle weights, with shaded areas representing the range of PSDs.
We observe that heavier vehicles tend to produce both quasi-static and surface wave signals with stronger amplitudes. 
These signals also exhibit higher signal-to-noise ratio and greater surface wave propagation distances, as evidenced in the example signals (Figures~\ref{fig:size}d-f). 
Additionally, surface waves induced by heavy vehicles have a more prominent mode in the low-frequency range (2.5--5 Hz) compared to lighter ones.

\par
\begin{figure}
    \centering
    \includegraphics[width=1\linewidth]{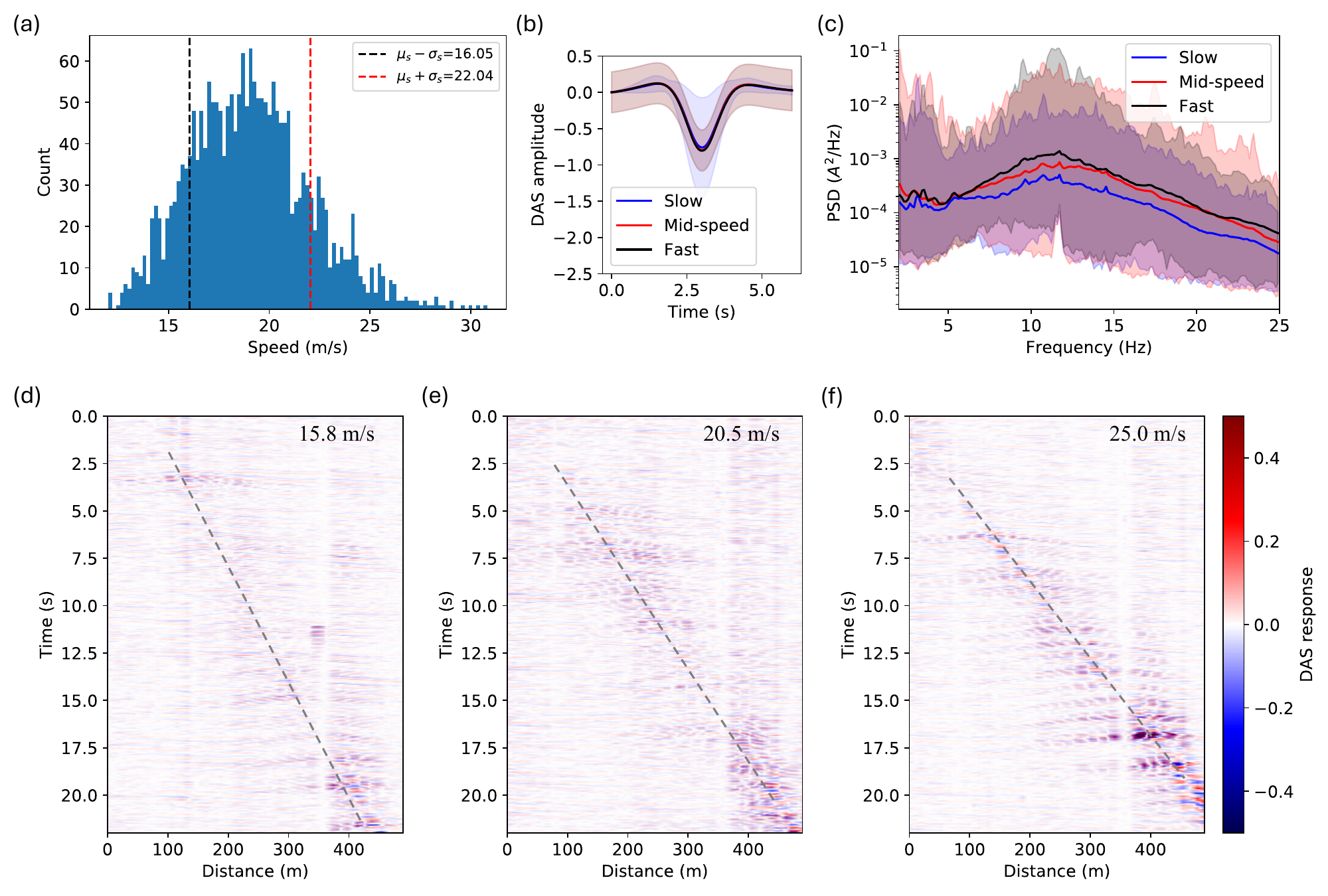}
    \caption{DAS signals induced by vehicles of different speeds. 
    (a) The distribution of vehicle speeds estimated using the specialized Kalman filter and smoothing algorithm. 
    Thresholds at the mean plus and minus the standard deviation of the vehicle speed differentiate between slow, mid-speed, and fast vehicles. 
    (b) The quasi-static DAS signals induced by vehicles of different speeds. 
    (c) The power spectral density (PSD) of surface wave signals for vehicles of different speeds. 
    PSD values at most frequencies slightly increase as the vehicle speed increases. 
    (d)-(f) Examples of surface wave signals for vehicles of different speeds.
    The dashed lines represent the estimated vehicle motions.
    The color version of this figure is available only in the electronic edition.}
    \label{fig:speed}
\end{figure}

We also investigate the DAS signals induced by vehicles traveling at different velocities, as illustrated in Figure~\ref{fig:speed}. 
Initially, to minimize the influence of vehicle weight and size, we select vehicles whose peak quasi-static signals are within one standard deviation of the mode peak value. 
{While this approach reduces the influence of vehicle weight and ensures a sufficient sample size for analysis, it does not fully disentangle the complex interplay between vehicle weight and velocity. Future research could leverage a more controlled dataset comprising vehicles with similar weights but varying speeds to refine our understanding of how velocity impacts vehicle-induced DAS signals.}
Subsequently, these vehicles are categorized into slow, mid-speed, and fast groups based on their average speeds, separated using one standard deviation ($\mu_s \pm \sigma_s$), as shown in Figure~\ref{fig:speed}a. 
There are 336 slow, 1,442 mid-speed, and 330 fast vehicles, respectively. 
Figures~\ref{fig:speed}d--f show the signal examples from vehicles traveling at different speeds with their traveling trajectories indicated by dashed lines, where a smaller slope corresponds to a faster speed.
Given the limited range of driving velocities, the surface wave signals among vehicles with varying speeds are not significantly different. 
Still, we note that surface waves induced by faster vehicles show slightly stronger amplitudes, which could be attributed to the greater kinetic energy and the corresponding larger impact forces exerted on rough road surfaces~\citep{jazar2024vehicle}.
Further research could explore the potential for more energetic surface wave signals generated by vehicles traveling at higher velocities, such as those on freeways, where pronounced vehicle-road interactions are likely due to increased velocity and the road roughness~\citep{yuan2023large}.

\subsection{Surface wave dispersion analysis}
\par
\begin{figure}
    \centering
    \includegraphics[width=1\linewidth]{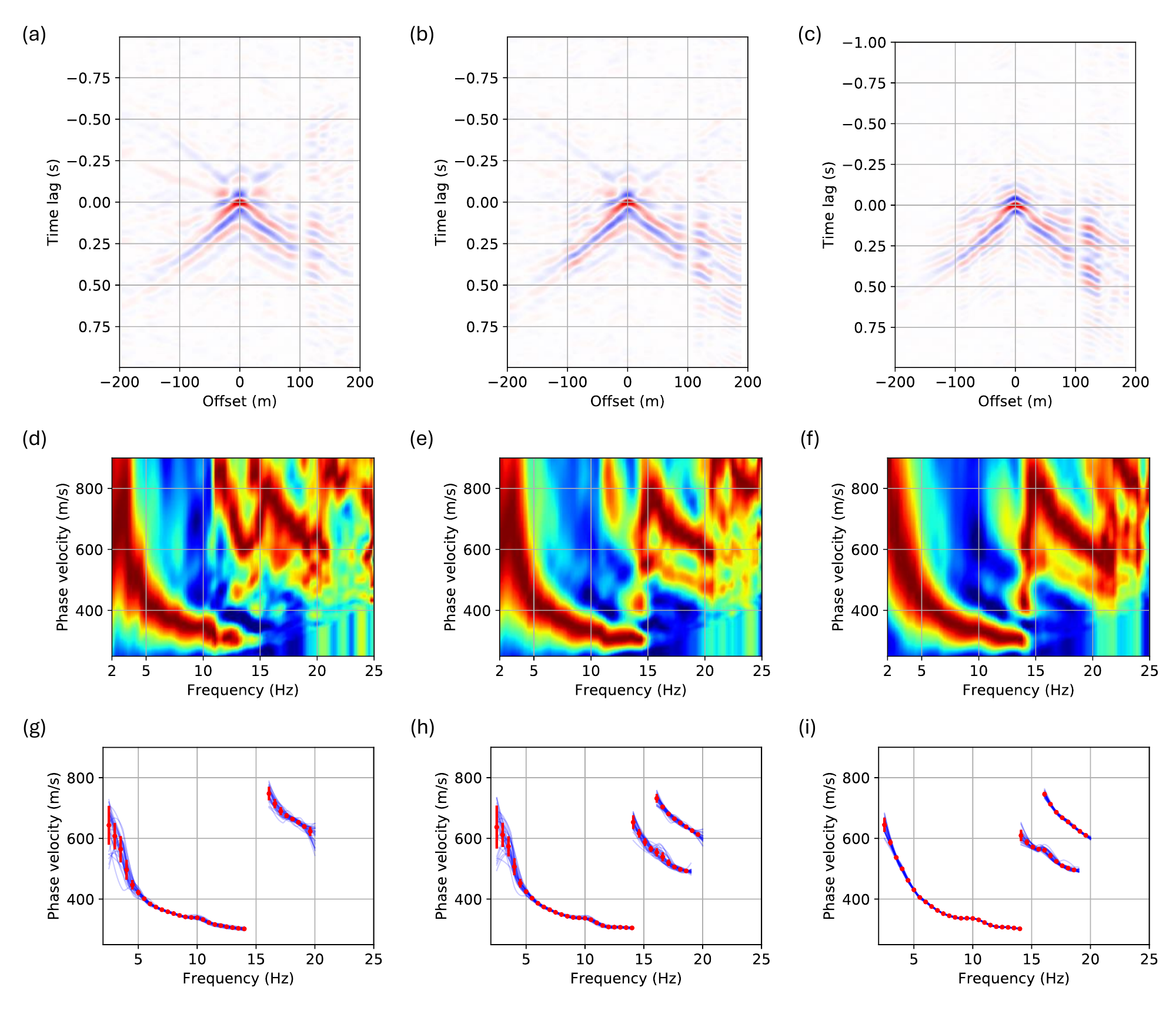}
    \caption{Retrieved VSGs and dispersion measurements from vehicles of different weights at the pivot location of 700 meters along the fiber (a grass lawn).
    (a)–(c) The VSGs stacked from the same number (60 VSGs) of passing vehicles with light, middle, and heavy weights, respectively. 
    (d)–(f) The computed dispersion spectra using the phase-shift method.
    (g)–(i) The picked dispersion curves from (d)–(f). 
    The blue lines show dispersion curves from VSGs constructed by stacking randomly sampled 60 vehicle-induced surface wave windows from each set of vehicles. 
    The red dots represent the averaged shear wave velocities, with error bars indicating the standard deviation.}
    \label{fig:dispersion_size}
\end{figure}

We use selected signal windows to measure the dispersion of surface waves for each vehicle class. 
We randomly sample and stack the same number of 60 VSGs to ensure a fair comparison for each class. 
We chose 60 VSGs because the resulting dispersion images become stable, as supported by our uncertainty quantification in the discussion section.
Figures~\ref{fig:dispersion_size}a--f show the retrieved VSGs at the pivot location of 700 meters along the fiber for light, mid-weight, and heavy vehicles, along with their respective dispersion spectra measurements. Additionally, for each vehicle class, we pick the dispersion curves 30 times from stacked VSGs derived from repeatedly sampled windows of 60 vehicle-induced surface waves each time. 
This process results in 30 dispersion curves for each vehicle class with the uncertain range, as shown in Figures~\ref{fig:dispersion_size}g–i.
\par
We find that the quality of both the VSGs and dispersion measurements improves with increasing vehicle weight. 
The retrieved VSGs from heavier vehicles (Figure~\ref{fig:dispersion_size}c) exhibit less noise in the negative time lag section, likely due to more accurate vehicle tracking and reduced ambient noise, as discussed in~\cite{https://doi.org/10.1029/2023JB028033}. 
{Heavier vehicles generate more pronounced quasi-static signals, facilitating more accurate vehicle detection and tracking, as well as higher signal-to-noise ratio surface wave windows. 
The stronger signals from the tracked heavy vehicles effectively overshadow the low-amplitude anti-causal signals in the VSGs, which are typically generated by vehicles traveling in distant opposing traffic lanes. This results in improved clarity and reliability of the VSGs and their corresponding dispersion spectra.}
On the dispersion spectra (Figures~\ref{fig:dispersion_size}d--f), the fundamental mode of dispersion shows greater continuity as vehicle weight increases, especially at low frequencies (<5 Hz).
From the picked dispersion curves for vehicles of different weights (Figures~\ref{fig:dispersion_size}g--i), we observe that those from heavy vehicles are more concentrated and show less uncertainty than those from light and mid-weight vehicles. 
Thus, low-frequency dispersion curves, which resolve deeper structures, can be more reliably measured from heavy vehicles.
Retrieving VSGs from heavy vehicles also improves the measurement of higher-order dispersion. 
Compared to the results from light vehicles (Figure~\ref{fig:dispersion_size}d), the higher mode dispersion curves are more distinguishable for mid-weight and heavy vehicles (Figures~\ref{fig:dispersion_size}e and f).
{Considering that light-to-middle-weight vehicles are approximately 17 times more than heavy vehicles during the same fixed time period, we retrieved VSGs by stacking 1,020 light-to-middle-weight vehicle signals and calculated the corresponding dispersion spectra for comparison, as shown in Figure S4. 
Despite stacking 1,020 VSGs, the resulting dispersion spectra for light-to-middle-weight vehicles still do not match the quality achieved with just 60 heavy vehicle VSGs.}
{When compared with ambient noise interferometry, }

\par
\begin{figure}
    \centering
    \includegraphics[width=1\linewidth]{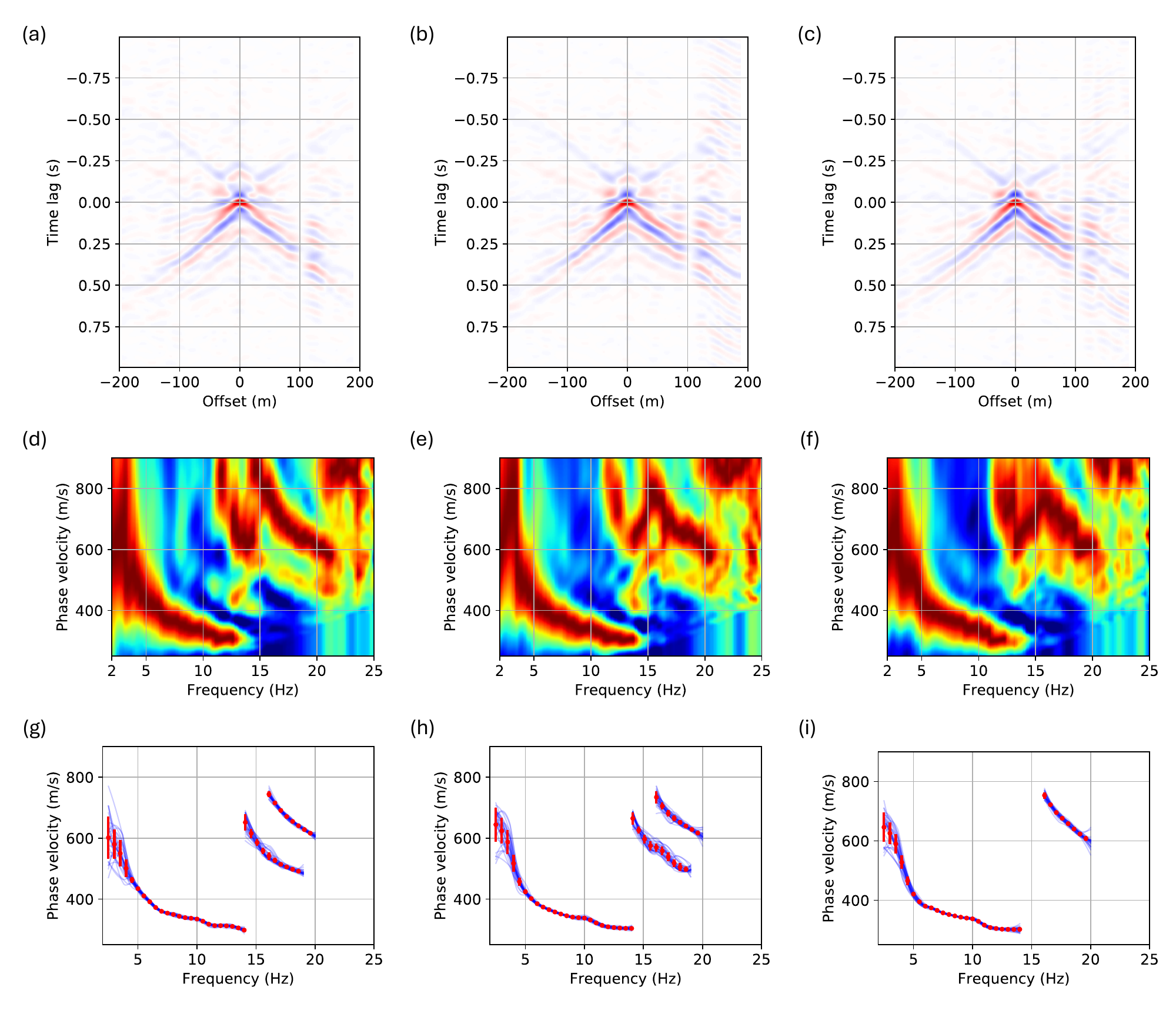}
    \caption{The same as Figure~\ref{fig:dispersion_size} but for vehicles of slow, middle, and fast traveling speeds.}
    \label{fig:dispersion_speed}
\end{figure}

For vehicles traveling at different speeds, we observe no significant differences in retrieved VSGs and measured dispersion curves, as shown in Figure~\ref{fig:dispersion_speed}. 
The low-frequency dispersion spectra for slow vehicles tend to be noisier than those for fast and mid-speed vehicles. 
Yet, only one high-order mode dispersion curve is consistently picked for fast vehicles.
The subtle variation observed here may not solely be attributed to vehicle speed, but other factors could also play a role, such as the high-cut filter effect imposed by the finite gauge length of the DAS interrogate~\citep{yuan2023large}.
Further investigation is needed to provide a more definitive explanation.
\par
We extended the dispersion analysis to different pivot locations, spaced 10 meters apart along the 500-700 meter section of the fiber. 
This analysis included various vehicle classes, and we consistently observed similar results across the different near-surface structures. 
Figures S1 and S2 showcase these results at the 600-meter pivot location, which corresponds to a built-up area different from the 700-meter pivot location. 
Although the VSGs and dispersion spectra vary between these two locations due to differences in subsurface structures, our findings confirm that the overall patterns in dispersion analysis across different vehicle classes remain consistent.

% \begin{figure*}
%     \centering
%     \includegraphics[width=1\linewidth]{figures/weight_disp_600.pdf}
%     \caption{The same as Fig~\ref{fig:dispersion_size} but at the pivot location of 600 meters along the fiber, which is a parking lot.
%     All three types of vehicles produce consistent fundamental mode Rayleigh surface waves.
%     The picked dispersion curves using heavy vehicles show the smallest variation compared to other types of vehicles.
%     }
%     \label{fig:dispersion_size_600}
% \end{figure*}

% \begin{figure*}
%     \centering
%     \includegraphics[width=1\linewidth]{figures/speed_disp_600.pdf}
%     \caption{The same as Fig~\ref{fig:dispersion_size_600} but for vehicles of different speeds.
%     The constructed VSGs and picked dispersion curves at this location with vehicles of different speeds are similar. 
%     }    \label{fig:dispersion_speed_600}
% \end{figure*}

\subsection{Inversion of shear-wave velocity}
\par
Beyond analyzing the features of the vehicle-induced DAS signals and the dispersion spectra, we inverted the dispersion curves to probe the subsurface structures.
We invert for parameters including thickness and shear-wave velocity ($V_s$) for each layer, assuming a six-layer model.
The density of each layer is approximated using an empirical relationship with the P-wave velocity as in~\cite{ji2024exploiting}. 
{From the VSGs, we estimate that the ratio between P-wave and S-wave velocities ranges from 2 to 4 at this location by measuring their travel time.
Based on the relationship between P-wave velocity, S-wave velocity, and Poisson's ratio, 
we calculated a Poisson's ratio range of 0.33 to 0.47, which was used in the inversion.}
The parameter ranges for inversion are specified in Table~\ref{tab:param}.
\par
\begin{figure}[!ht]
    \centering
    \includegraphics[width=1\linewidth]{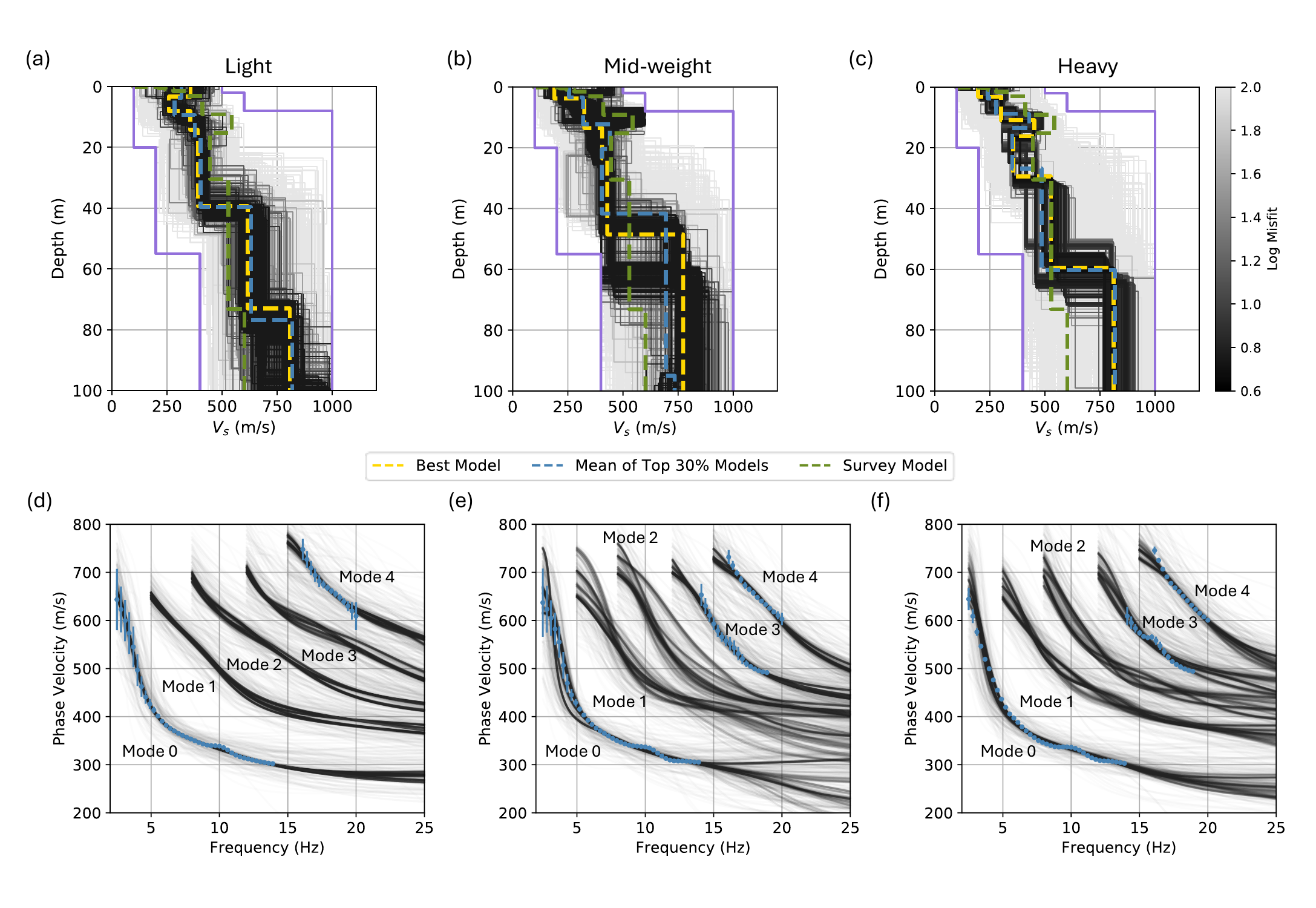}
    \caption{
    Inversion results of near-surface structures at the 700-meter pivot location. 
    (a)-(c) 1-D $V_S$ model of the near-surface structure for light, mid-weight, and heavy vehicles, respectively. 
    Gray colors indicate the logarithmic misfits. 
    Purple solid lines show the model parameter boundaries detailed in Table~\ref{tab:param}. 
    Yellow and blue dashed lines represent the best model (detailed in Table~\ref{tab:best}) with the lowest misfit value and the mean of the top 30\% model. The green dashed line shows the velocity model from the SASW survey.
    (d)-(f) Model fit of dispersion curves for light, mid-weight, and heavy vehicles, respectively. 
    Blue dots with error bars are estimations from Figure~\ref{fig:dispersion_size}. 
    Lines are model predictions of dispersion curves, including the top 70\% of the predicted models. 
    The color version of this figure is available only in the electronic edition.
     }
    \label{fig:imaging_weights}
\end{figure}
\begin{figure}[!ht]
    \centering
    \includegraphics[width=1\linewidth]{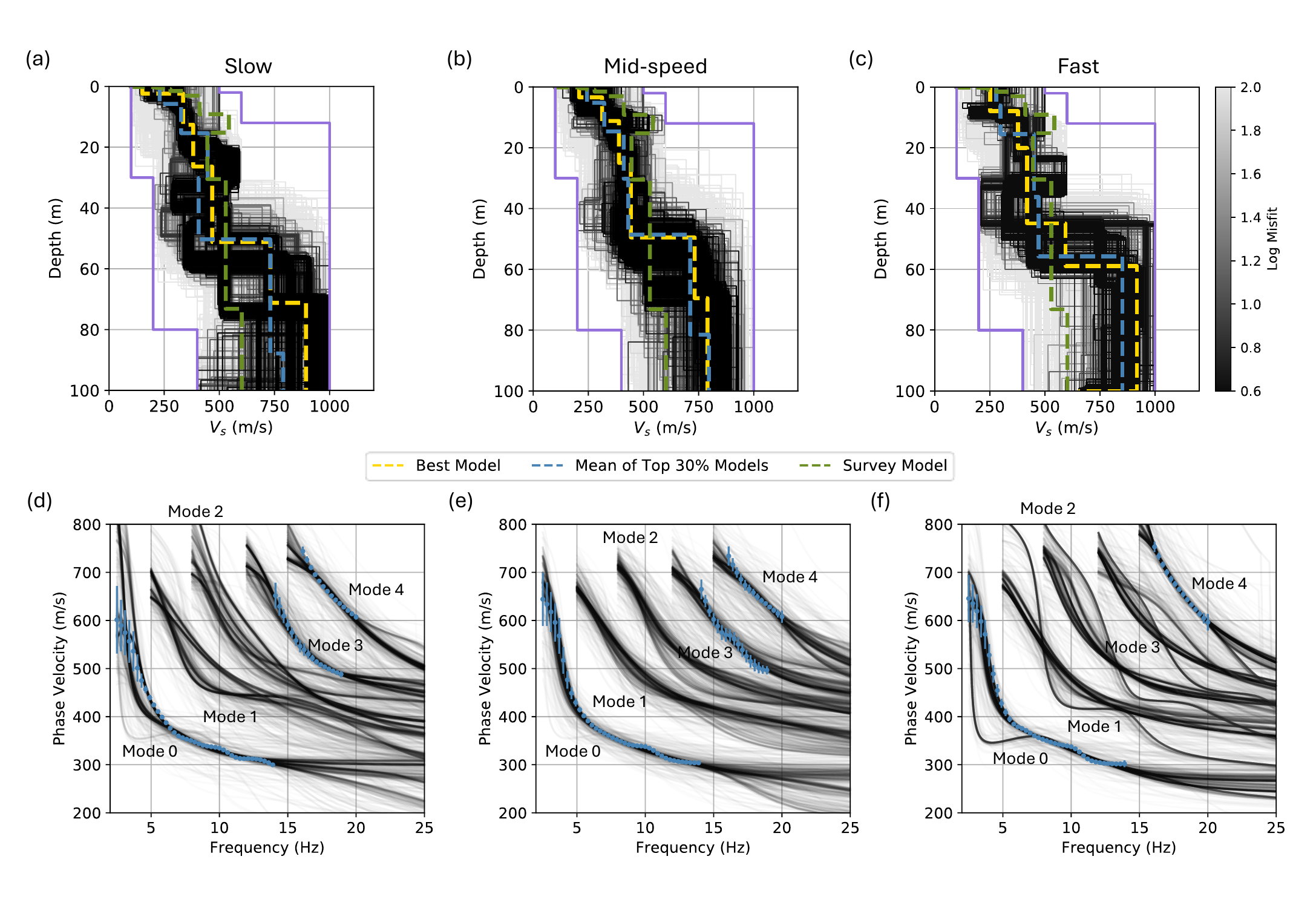}
    \caption{The same as Figure~\ref{fig:imaging_weights} but for vehicles of slow, middle, and fast traveling speeds. The color version of this figure is available only in the electronic edition.
     }
    \label{fig:imaging_speeds}
\end{figure}

We compare the inversion of shear-wave velocity with different vehicle classes at the 700-meter pivot location. 
The mean values and ranges of the dispersion curves, obtained through the 30 times dispersion curve picking, are input into the CPSO inversion workflow. 
After running the CPSO inversion five times with a swarm size of 50 and a maximum of 1000 iterations, we obtain the inversion results for near-surface structures.
The best models and the top 30\% of models are shown in the first row of Figures~\ref{fig:imaging_weights} and~\ref{fig:imaging_speeds} to demonstrate model convergence toward the measurements. 
The second row of Figures~\ref{fig:imaging_weights} and~\ref{fig:imaging_speeds} presents the model fit of dispersion curves, overlaid with our estimates from the dispersion spectra.
\par
Our inversion results using heavy vehicles align most closely with the 2009 SASW results~\citep{Wong_Stokoe_2009}, especially at the depth range of 30 to 60 meters, compared with the results from light and mid-weight vehicles.
Table~\ref{tab:best} presents the inverted parameters using surface waves generated by heavy vehicles.
Using heavy vehicles allows us to pick the fundamental mode dispersion curves at a low frequency more robustly, which improves the imaging of structures at greater depth.
It is noteworthy that the SASW model, with a frequency range from approximately 3 Hz to over 100 Hz, resolves the uppermost 10 meters of the subsurface with greater detail than our results.
This resolution advantage is attributed to the higher frequency range of the SASW data, which enhances the detection of finer details at shallower depths. 
Our inversion result also differs from the SASW survey for structures deeper than 60 meters. 
This discrepancy may be due to the small sensitivity in the frequency range of our estimated dispersion curves, detailed in the discussion section.
We also speculate, without being able to prove it, that our results benefit from the high signal-to-noise ratio achieved at low frequency by analyzing surface waves generated by heavy vehicles.
\begin{table}[t!]
\centering
\caption{Parameter~ranges~for~inversion.}
\begin{tabular}{cccc}
\toprule
Layer & Thickness (m) & Vs (m/s)\\
\midrule
1 & 1 -- 10 & 100 -- 500\\
2 & 1 -- 10 & 100 -- 500\\
3 & 1 -- 10 & 200 -- 600\\
4 & 5 -- 25 & 200 -- 600\\
5 & 20 -- 80 & 400 -- 1000\\
Half-space & -- & 400 -- 1000\\
\bottomrule
\label{tab:param}
\end{tabular}
\end{table}

\begin{table}[t!]
\centering
\caption{Inversion~results~for~the~best~model.}
\begin{tabular}{cccc}
\toprule
Layer & Thickness (m) & Vs (m/s) \\
\midrule
1 & 3 & 198 \\
2 & 8 & 302 \\
3 & 5 & 452 \\
4 & 13 & 357 \\
5 & 30 & 527 \\
Half-space & -- & 811\\
\bottomrule
\label{tab:best}
\end{tabular}
\end{table}

\section{Discussion}
\subsection{Uncertainty quantification of dispersion curves}
\par
Since our approach stacks multiple VSGs from vehicle-induced surface wave windows to improve the signal-to-noise ratio, it is important to examine the convergence performance of the uncertainties in the picked dispersion curves as the number of stacked VSGs increases. We employ the bootstrapping method~\citep{mooney1993bootstrapping} to quantify the uncertainty of dispersion curves by calculating their standard deviations. 
The details of the bootstrapping process are as follows:
\begin{itemize}
    \item[1.] Randomly sample $N$ vehicle-induced surface wave windows with replacement from all available surface wave windows at one location for the same vehicle class.
    \item[2.] Compute the dispersion spectra of the stacked VSGs constructed from these $N$ surface wave windows.
    \item[3.] Pick the dispersion curve on the dispersion spectra by peak detection.
    \item[4.] Repeat steps 1--3 for $M$ times, resulting in $M$ dispersion curves of each mode.
    \item[5.] Calculate the standard deviation $\sigma_N$ of the $M$ dispersion curves.
\end{itemize}
Here, $\sigma_N$ quantifies the uncertainty of the dispersion curves computed using VSGs constructed from $N$ vehicle-induced surface wave windows. 
\par
\begin{figure}[ht!]
    \centering
    \includegraphics[width=1\linewidth]{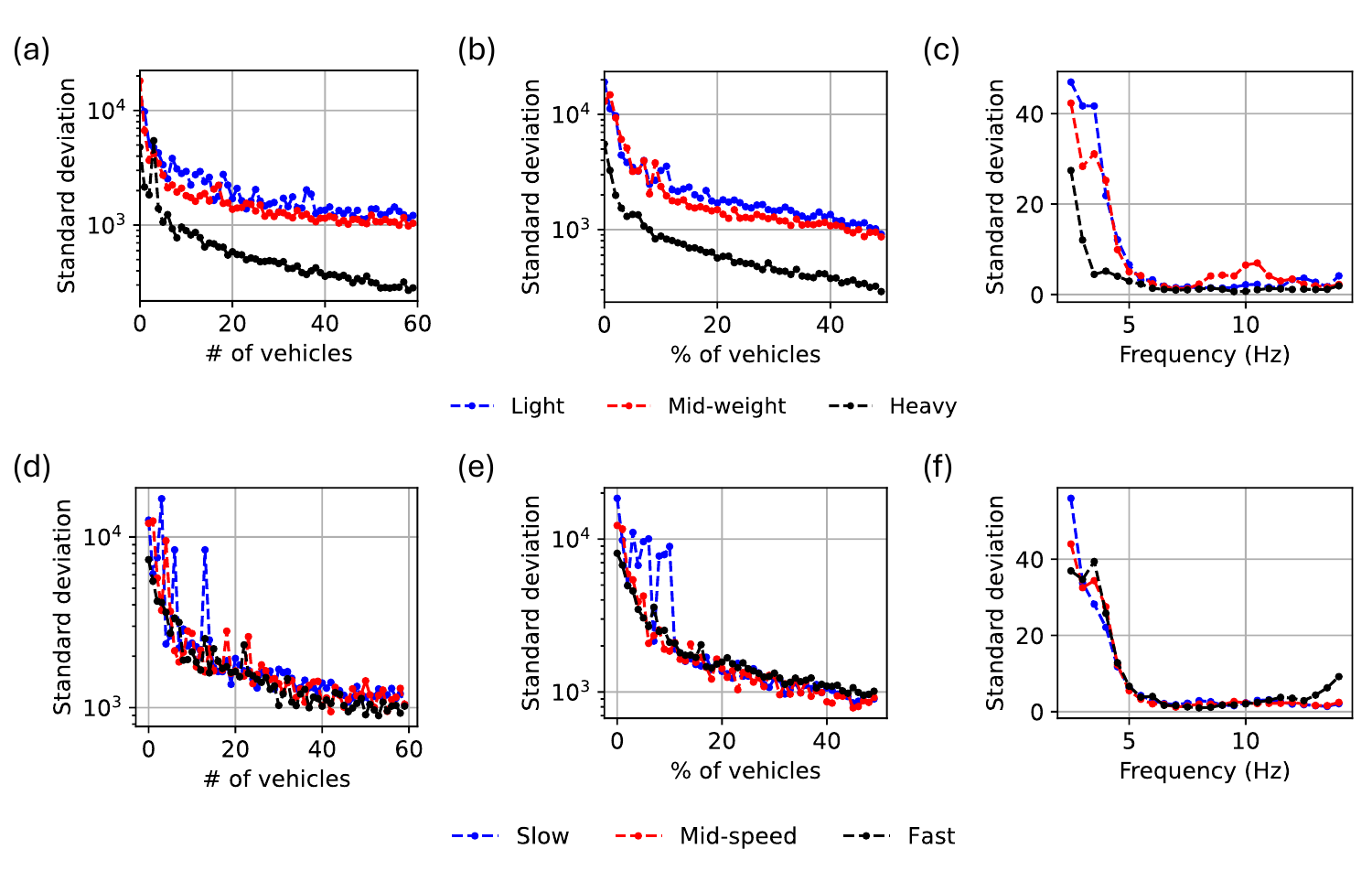}
    \caption{
    Uncertainty quantification of the picked dispersion curves for different numbers of vehicles with varying weights and speeds. 
    The uncertainties are quantified by bootstrapping the standard deviations of the estimated fundamental mode dispersion curves. (a) shows the standard deviations of the fundamental mode dispersion curves for vehicles with different weights at pivot locations of 700 meters.
    (b) shows the standard deviations of the fundamental mode dispersion curves based on the percentage of vehicles' VSGs stacked with varying vehicle weights.
    (c) displays the converged standard deviations of the fundamental mode dispersion curves at various frequencies for vehicles with different weights.
    (d), (e) and (f) present the same data as (a), (b), and (c) but for vehicles with different speeds. 
    The color version of this figure is available only in the electronic edition.
    } 
    \label{fig:convergence_size}
\end{figure}

We gradually increase $N$, the number of VSGs stacked, to study the convergence of the uncertainty of the fundamental mode dispersion curves {(2.5 to 14 Hz)} and to estimate how many vehicles are needed to achieve convergence in dispersion curve picking.
For each bootstrapping, M is set to 30 to ensure sufficient samples for calculating the standard deviation while maintaining manageable data volumes for processing.
We apply the bootstrapping method at both pivot locations of 600 and 700 meters.
Figures~\ref{fig:convergence_size}a and d show that the uncertainties (standard deviation) converged after stacking approximately 55 VSGs across all cases. 
The 600-meter pivot location shows a similar result, as detailed in the Supplemental Materials (Figure S3).
The converged standard deviation of dispersion curves for heavy vehicles is about four times smaller than that for light and mid-weight vehicles at the 700-meter pivot location and five times smaller at the 600-meter location. 
The converged standard deviations of dispersion curves for vehicles with different speeds are similar.
However, stacking VSGs from fewer slow-moving vehicles (e.g., fewer than 10 vehicles) results in larger uncertainty in dispersion curves.
{We also sampled an equal number of surface wave windows from each vehicle class to create subsets with the same number of observations. 
Subsequently, we perform an additional uncertainty quantification of the picked dispersion curves based on the percentage of vehicles relative to the total number of vehicles within each subset, as shown in Figures~\ref{fig:convergence_size}b and e. 
In particular, we calculate the standard deviation of the picked dispersion curves while incrementally increasing the percentage of VSGs stacked.
This approach ensures a fair comparison of uncertainty, accounting for the differing numbers of recorded vehicles across weight classes during the fixed data collection period.
Consistent with previous observations, stacking the same percentage of vehicles reveals that the low-frequency dispersion curves for heavy vehicles exhibit lower uncertainty (standard deviation) compared to those for light and mid-weight vehicles.
The percentage of stacked vehicles is capped at 50\% due to limitations in the bootstrapping algorithm. 
The algorithm estimates uncertainty by randomly sampling vehicle-induced surface wave windows with replacement from all available windows within the same vehicle class, repeating this process multiple times. 
A very large sampling size introduces a significant issue: the limited population size leads to a high probability of repeatedly sampling the same vehicles. This reduces the variability and independence of the subsamples, undermining the statistical robustness and accuracy of the bootstrapped estimates.
}

\par
We further investigate the converged uncertainty across different frequencies. 
Figure~\ref{fig:convergence_size}c and d display the converged standard deviations of the fundamental mode phase velocities at various frequencies for different vehicle classes.
At frequencies below 5 Hz, the converged uncertainties for heavy vehicles are significantly smaller than those for light and mid-weight vehicles. 
For example, at 3 Hz, the converged uncertainty for heavy vehicles is approximately 3$\times$ smaller than that for light and mid-weight vehicles at the 700-meter pivot location. 
Another pivot location at 600 meters presents similar results, as shown in the Supplemental Materials (Figure S3).
At higher frequencies, the converged uncertainties of phase velocities for vehicles of different weights are similar. 
Moreover, the converged uncertainties of phase velocities at different frequencies remain consistent across vehicles traveling at different speeds.

\subsection{Surface wave sensitivity analysis}
\begin{figure}[!ht]
    \centering
    \includegraphics[width=0.5\linewidth]{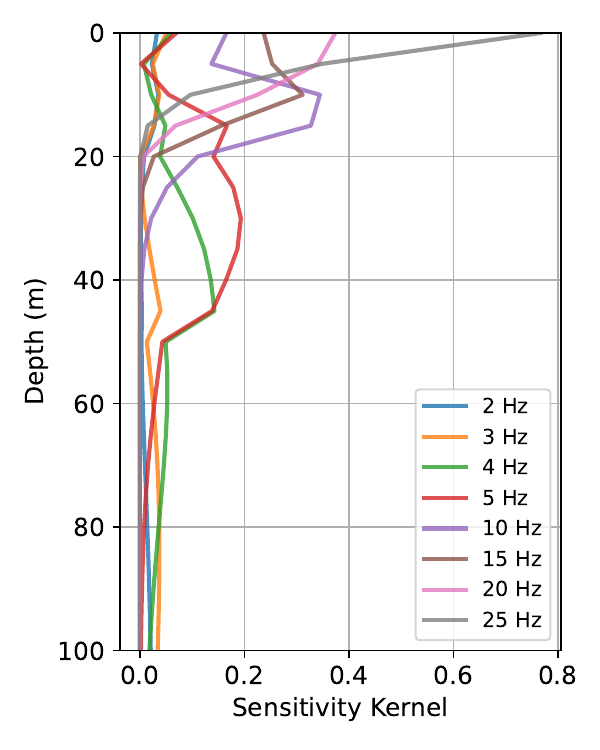}
    \caption{Rayleigh surface‐wave phase velocity sensitivity kernels, corresponding to the best model derived from heavy vehicle signals (yellow dashed line in Figure~\ref{fig:imaging_weights}c). 
    The color version of this figure is available only in the electronic edition.
    } 
    \label{fig:sensitivity}
\end{figure}

The depth sensitivity of surface waves is closely related to their frequency content.
The Rayleigh surface-wave phase velocity sensitivity kernels, corresponding to the best model derived from heavy vehicle signals {(yellow dashed line in Figure~\ref{fig:imaging_weights}c)}, are shown in Figure~\ref{fig:sensitivity}.
It can be observed that surface waves at lower frequencies (2--5 Hz) are primarily influenced by structures between 30 to 80 meters in depth. 
Improved estimation of phase velocities at these lower frequencies enhances sensitivity for imaging deeper structures, whereas structures above the uppermost 10 meters are more sensitive at frequencies above 10 Hz.

\subsection{Insights into near-surface characterization using vehicle-induced DAS signals}
Our analysis and results provide insights into judiciously selecting vehicle-induced DAS signals to achieve more accurate and efficient dispersion curve picking and near-surface imaging. 
Previous research introduced continuous near-surface characterization using these signals, employing all isolated vehicles regardless of their characteristics to construct VSGs and perform dispersion analysis~\citep{https://doi.org/10.1029/2023JB028033}.
Building on this foundation, our work demonstrates that stacking VSGs reduces the uncertainty of picked dispersion curves, converging after signals from approximately 55 vehicles' VSGs are stacked. 
Converged VSGs and dispersion curves obtained using DAS signals from heavy vehicles exhibit a higher signal-to-noise ratio at frequencies below 5 Hz compared to those from lighter vehicles, allowing for better constraints of subsurface structures at greater depths.
Using lighter vehicles shows comparable performance at frequencies above 5 Hz after convergence. 
Additionally, there are no significant differences in the VSGs and dispersion curves obtained using vehicles traveling at different speeds.
\par
Although the dispersion measurement varies with vehicle characteristics, the frequency of occurrence of different vehicle classes also varies. 
For instance, we can only obtain an average of 112 isolated surface wave windows from heavy vehicles over 15 days, equating to about seven windows per day. 
In contrast, light and mid-weight vehicles are more common, with an average of 119 isolated vehicles recorded per day. 
This abundance allows for daily converged VSGs and dispersion curves using light and mid-weight vehicles. 
However, achieving converged VSGs and dispersion curves from heavy vehicles would require about eight days of recordings.
Notably, at frequencies below 5 Hz, even stacking VSGs from ten heavy vehicles results in lower uncertainty than converged VSGs using only lighter vehicles. 
Additionally, the frequency of vehicle occurrence with different weights can vary depending on the road's function. 
The relatively low number of heavy vehicles in our study may be attributed to the fact that the road we analyzed is an arterial road, which typically sees a low volume of heavy vehicles per day.
Based on our findings, we have the following recommendations:
\begin{itemize}
    \item[1. ] For fine-grained temporal monitoring of near-surface structures (e.g., daily monitoring), the high-quality dispersion curve can be estimated piece-wise: low frequencies (e.g., $<$ 5 Hz) should be derived from stacking as many heavy vehicle VSGs as possible, while high frequencies can be derived from stacking VSGs from all vehicles until convergence is achieved.
    \item[2. ] For coarse-grained temporal monitoring of near-surface structures (e.g., weekly monitoring), the high-quality dispersion curve should be estimated from stacked heavy vehicle VSGs, which is more accurate and efficient as only a few surface wave windows are required for cross-correlation.
\end{itemize}
The number of vehicles needed to achieve converged dispersion curves may vary for different near-surface structures and fiber coupling methods. 
Thus, we suggest conducting uncertainty quantification first to determine the needed number for achieving convergence and choose the appropriate method according to the required temporal resolution of near-surface monitoring.

\section{Conclusion}
\par
This study characterizes surface wave data generated by moving vehicles of different weights/sizes and speeds in the Stanford DAS-2 experiment, offering insights into accurate and efficient near-surface characterization using vehicle-induced DAS data.
Our findings reveal that vehicle weight and size significantly influence the surface wave signals and, consequently, the inversion results.
Specifically, heavy-weight vehicles generate higher signal-to-noise ratios and reduced uncertainty in phase velocity estimates, particularly at lower frequencies (<5 Hz), which are crucial for probing deeper near-surface structures. 
The impact of vehicle speed on the quality of the dispersion images is found to be less pronounced than that of vehicle weight.
\par
Our study suggests that judiciously selecting and processing signals from specific vehicle types can enhance near-surface imaging quality. 
For fine-grained temporal monitoring, low-frequency dispersion curves can be derived from heavy-weight vehicle signals, whereas high-frequency curves can be obtained from all vehicle classes until convergence is achieved. 
Utilizing signals from heavy-weight vehicles alone is efficient and accurate for coarse-grained temporal monitoring.
Our results and insights underscore the potential of DAS technology in continuous urban seismic monitoring and the importance of considering vehicle characteristics to optimize data acquisition and processing. 

\section{Data and Resources}
\par
The data supporting this study are derived from continuous DAS recordings captured by the Stanford DAS-2 array.
The surface wave data at the pivot locations of 600 and 700 meters is publicly available at \url{https://doi.org/10.5281/zenodo.12775602}.
The code implementation is also publicly available at \url{https://github.com/jingxiaoliu/das_diff_veh}.

\section{Declaration of Competing Interests}
\par
The authors acknowledge there are no conflicts of interest recorded.

\begin{ack}
The interrogator unit was loaned to us by Luna--OptaSense. We thank Martin Karrenbach, Victor Yartsev, and Lisa LaFlame from Optasense, as well as the Stanford ITS fiber team, and in particular Erich Snow, for crucial help with the Stanford DAS-2 experiment. We also thank the Stanford School of Earth IT team for hosting the interrogator in the computer room. We thank Bin Luo and Qing Ji for the ambient noise interferometry and inversion codes. 
\end{ack}

\bibliography{citation.bib}

\begin{thebibliography}{}

\bibitem[\protect\citeauthoryear{Ajo-Franklin, Dou, Lindsey, Monga, Tracy, Robertson, Rodriguez~Tribaldos, Ulrich, Freifeld, Daley, and Li}{Ajo-Franklin et~al.}{2019}]{Ajo-Franklin2019}
Ajo-Franklin, J.~B., S.~Dou, N.~J. Lindsey, I.~Monga, C.~Tracy, M.~Robertson, V.~Rodriguez~Tribaldos, C.~Ulrich, B.~Freifeld, T.~Daley, and X.~Li (2019).
\newblock Distributed acoustic sensing using dark fiber for near-surface characterization and broadband seismic event detection.
\newblock {\em Scientific Reports\/}~{\bf 9\/}(1), 1328.

\bibitem[\protect\citeauthoryear{Dasios, McCann, Astin, McCann, and Fenning}{Dasios et~al.}{1999}]{https://doi.org/10.1046/j.1365-2478.1999.00138.x}
Dasios, McCann, Astin, McCann, and Fenning (1999).
\newblock Seismic imaging of the shallow subsurface: shear-wave case histories.
\newblock {\em Geophysical Prospecting\/}~{\bf 47\/}(4), 565--591.

\bibitem[\protect\citeauthoryear{Dou, Lindsey, Wagner, Daley, Freifeld, Robertson, Peterson, Ulrich, Martin, and Ajo-Franklin}{Dou et~al.}{2017}]{Dou2017}
Dou, S., N.~Lindsey, A.~M. Wagner, T.~M. Daley, B.~Freifeld, M.~Robertson, J.~Peterson, C.~Ulrich, E.~R. Martin, and J.~B. Ajo-Franklin (2017).
\newblock Distributed acoustic sensing for seismic monitoring of the near surface: A traffic-noise interferometry case study.
\newblock {\em Scientific Reports\/}~{\bf 7\/}(1), 11620.

\bibitem[\protect\citeauthoryear{Eberhart and Shi}{Eberhart and Shi}{2000}]{870279}
Eberhart, R. and Y.~Shi (2000).
\newblock Comparing inertia weights and constriction factors in particle swarm optimization.
\newblock In {\em Proceedings of the 2000 Congress on Evolutionary Computation. CEC00 (Cat. No.00TH8512)}, Volume~1, pp.\  84--88 vol.1.

\bibitem[\protect\citeauthoryear{Fang, Li, Zhao, and Martin}{Fang et~al.}{2020}]{https://doi.org/10.1029/2019GL086115}
Fang, G., Y.~E. Li, Y.~Zhao, and E.~R. Martin (2020).
\newblock Urban near-surface seismic monitoring using distributed acoustic sensing.
\newblock {\em Geophysical Research Letters\/}~{\bf 47\/}(6), e2019GL086115.
\newblock e2019GL086115 10.1029/2019GL086115.

\bibitem[\protect\citeauthoryear{Ganji, Gucunski, and Nazarian}{Ganji et~al.}{1998}]{doi:10.1061/(ASCE)1090-0241(1998)124:8(757)}
Ganji, V., N.~Gucunski, and S.~Nazarian (1998).
\newblock Automated inversion procedure for spectral analysis of surface waves.
\newblock {\em Journal of Geotechnical and Geoenvironmental Engineering\/}~{\bf 124\/}(8), 757--770.

\bibitem[\protect\citeauthoryear{Jazar and Marzbani}{Jazar and Marzbani}{2024}]{jazar2024vehicle}
Jazar, R.~N. and H.~Marzbani (2024).
\newblock {\em Vehicle Vibrations: Linear and Nonlinear Analysis, Optimization, and Design}.
\newblock Springer Nature.

\bibitem[\protect\citeauthoryear{Ji, Luo, and Biondi}{Ji et~al.}{2024a}]{ji2024exploiting}
Ji, Q., B.~Luo, and B.~Biondi (2024a).
\newblock Exploiting the potential of urban das grids: Ambient-noise subsurface imaging using joint rayleigh and love waves.
\newblock {\em Seismological Research Letters\/}~{\bf 95\/}(3), 1794--1811.

\bibitem[\protect\citeauthoryear{Ji, Luo, and Biondi}{Ji et~al.}{2024b}]{10.1785/0220230104}
Ji, Q., B.~Luo, and B.~Biondi (2024b, 01).
\newblock {Exploiting the Potential of Urban DAS Grids: Ambient‐Noise Subsurface Imaging Using Joint Rayleigh and Love Waves}.
\newblock {\em Seismological Research Letters\/}~{\bf 95\/}(3), 1794--1811.

\bibitem[\protect\citeauthoryear{Lei and Wang}{Lei and Wang}{2024}]{10.1785/0220240050}
Lei, Y. and B.~Wang (2024, 05).
\newblock {Illuminating Urban Near‐Surface with Distributed Acoustic Sensing Multimodal Noise Surface‐Wave Imaging}.
\newblock {\em Seismological Research Letters\/}.

\bibitem[\protect\citeauthoryear{Lindsey and Martin}{Lindsey and Martin}{2021}]{annurev:/content/journals/10.1146/annurev-earth-072420-065213}
Lindsey, N.~J. and E.~R. Martin (2021).
\newblock Fiber-optic seismology.
\newblock {\em Annual Review of Earth and Planetary Sciences\/}~{\bf 49\/}(Volume 49, 2021), 309--336.

\bibitem[\protect\citeauthoryear{Liu, Yuan, Dong, Biondi, and Noh}{Liu et~al.}{2023}]{10.1145/3596262}
Liu, J., S.~Yuan, Y.~Dong, B.~Biondi, and H.~Y. Noh (2023, jun).
\newblock Telecomtm: A fine-grained and ubiquitous traffic monitoring system using pre-existing telecommunication fiber-optic cables as sensors.
\newblock {\em Proc. ACM Interact. Mob. Wearable Ubiquitous Technol.\/}~{\bf 7\/}(2).

\bibitem[\protect\citeauthoryear{Luu}{Luu}{2023}]{luu_2023_8007868}
Luu, K. (2023, June).
\newblock {evodcinv: Inversion of dispersion curves using Evolutionary Algorithms}.
\newblock {\em Zenodo\/}.

\bibitem[\protect\citeauthoryear{Martin, Biondi, Ajo-Franklin, and Papanicolaou}{Martin et~al.}{2018}]{2437101751}
Martin, E.~R., B.~Biondi, J.~Ajo-Franklin, and G.~Papanicolaou (2018).
\newblock Passive imaging and characterization of the subsurface with distributed acoustic sensing.
\newblock {\em ProQuest Dissertations and Theses\/}.

\bibitem[\protect\citeauthoryear{Martin, Castillo, Cole, Sawasdee, Yuan, Clapp, Karrenbach, and Biondi}{Martin et~al.}{2017}]{10.1190/tle36121025.1}
Martin, E.~R., C.~M. Castillo, S.~Cole, P.~S. Sawasdee, S.~Yuan, R.~Clapp, M.~Karrenbach, and B.~L. Biondi (2017, 12).
\newblock {Seismic monitoring leveraging existing telecom infrastructure at the SDASA: Active, passive, and ambient-noise analysis}.
\newblock {\em The Leading Edge\/}~{\bf 36\/}(12), 1025--1031.

\bibitem[\protect\citeauthoryear{Masoudi and Newson}{Masoudi and Newson}{2016}]{10.1063/1.4939482}
Masoudi, A. and T.~P. Newson (2016, 01).
\newblock {Contributed Review: Distributed optical fibre dynamic strain sensing}.
\newblock {\em Review of Scientific Instruments\/}~{\bf 87\/}(1), 011501.

\bibitem[\protect\citeauthoryear{Mooney, Duval, and Duvall}{Mooney et~al.}{1993}]{mooney1993bootstrapping}
Mooney, C.~Z., R.~D. Duval, and R.~Duvall (1993).
\newblock {\em Bootstrapping: A nonparametric approach to statistical inference}.
\newblock Number~95. sage.

\bibitem[\protect\citeauthoryear{Park, Miller, and Xia}{Park et~al.}{1999}]{doi:10.1190/1.1444590}
Park, C.~B., R.~D. Miller, and J.~Xia (1999).
\newblock Multichannel analysis of surface waves.
\newblock {\em GEOPHYSICS\/}~{\bf 64\/}(3), 800--808.

\bibitem[\protect\citeauthoryear{Park, Miller, and Xia}{Park et~al.}{2005}]{doi:10.1190/1.1820161}
Park, C.~B., R.~D. Miller, and J.~Xia (2005).
\newblock {\em Imaging dispersion curves of surface waves on multi‐channel record}, pp.\  1377--1380.

\bibitem[\protect\citeauthoryear{Rauch, Tung, and Striebel}{Rauch et~al.}{1965}]{doi:10.2514/3.3166}
Rauch, H.~E., F.~Tung, and C.~T. Striebel (1965).
\newblock Maximum likelihood estimates of linear dynamic systems.
\newblock {\em AIAA Journal\/}~{\bf 3\/}(8), 1445--1450.

\bibitem[\protect\citeauthoryear{Shapiro, Campillo, Stehly, and Ritzwoller}{Shapiro et~al.}{2005}]{doi:10.1126/science.1108339}
Shapiro, N.~M., M.~Campillo, L.~Stehly, and M.~H. Ritzwoller (2005).
\newblock High-resolution surface-wave tomography from ambient seismic noise.
\newblock {\em Science\/}~{\bf 307\/}(5715), 1615--1618.

\bibitem[\protect\citeauthoryear{van~den Bergh and Engelbrecht}{van~den Bergh and Engelbrecht}{2001}]{2890425708}
van~den Bergh, F. and A.~P. Engelbrecht (2001).
\newblock An analysis of particle swarm optimizers.
\newblock {\em PQDT - Global\/}, 301.

\bibitem[\protect\citeauthoryear{Wapenaar, Draganov, Snieder, Campman, and Verdel}{Wapenaar et~al.}{2010}]{doi:10.1190/1.3457445}
Wapenaar, K., D.~Draganov, R.~Snieder, X.~Campman, and A.~Verdel (2010).
\newblock Tutorial on seismic interferometry: Part 1 — basic principles and applications.
\newblock {\em GEOPHYSICS\/}~{\bf 75\/}(5), 75A195--75A209.

\bibitem[\protect\citeauthoryear{Wong and Stokoe}{Wong and Stokoe}{2009}]{Wong_Stokoe_2009}
Wong, I. and K.~H. Stokoe (2009, Aug).
\newblock {SASW measurements at Stanford University}.

\bibitem[\protect\citeauthoryear{Yang, Atterholt, Shen, Muir, Williams, and Zhan}{Yang et~al.}{2022}]{https://doi.org/10.1029/2021GL096503}
Yang, Y., J.~W. Atterholt, Z.~Shen, J.~B. Muir, E.~F. Williams, and Z.~Zhan (2022).
\newblock Sub-kilometer correlation between near-surface structure and ground motion measured with distributed acoustic sensing.
\newblock {\em Geophysical Research Letters\/}~{\bf 49\/}(1), e2021GL096503.
\newblock e2021GL096503 2021GL096503.

\bibitem[\protect\citeauthoryear{Yuan}{Yuan}{2023}]{yuan2023large}
Yuan, S. (2023).
\newblock {\em Large-Scale and Continuous Subsurface Monitoring Using Distributed Acoustic Sensing in Urban Environments}.
\newblock Ph.\ D. thesis, Stanford University.

\bibitem[\protect\citeauthoryear{Yuan, Lellouch, Clapp, and Biondi}{Yuan et~al.}{2020}]{10.1190/tle39090646.1}
Yuan, S., A.~Lellouch, R.~G. Clapp, and B.~Biondi (2020, 09).
\newblock {Near-surface characterization using a roadside distributed acoustic sensing array}.
\newblock {\em The Leading Edge\/}~{\bf 39\/}(9), 646--653.

\bibitem[\protect\citeauthoryear{Yuan, Liu, Noh, Clapp, and Biondi}{Yuan et~al.}{2024}]{https://doi.org/10.1029/2023JB028033}
Yuan, S., J.~Liu, H.~Y. Noh, R.~Clapp, and B.~Biondi (2024).
\newblock Using vehicle-induced das signals for near-surface characterization with high spatiotemporal resolution.
\newblock {\em Journal of Geophysical Research: Solid Earth\/}~{\bf 129\/}(4), e2023JB028033.
\newblock e2023JB028033 2023JB028033.

\bibitem[\protect\citeauthoryear{Yuan, Liu, Young~Noh, and Biondi}{Yuan et~al.}{2021}]{10.1190/segam2021-3584136.1}
Yuan, S., J.~Liu, H.~Young~Noh, and B.~Biondi (2021, 09).
\newblock {Urban system monitoring using combined vehicle onboard sensing and roadside distributed acoustic sensing}.
\newblock Volume Day 1 Sun, September 26, 2021 of {\em SEG International Exposition and Annual Meeting}, pp.\  D011S137R003.

\bibitem[\protect\citeauthoryear{Yuan, van~den Ende, Liu, Noh, Clapp, Richard, and Biondi}{Yuan et~al.}{2024}]{10280620}
Yuan, S., M.~van~den Ende, J.~Liu, H.~Y. Noh, R.~Clapp, C.~Richard, and B.~Biondi (2024).
\newblock Spatial deep deconvolution u-net for traffic analyses with distributed acoustic sensing.
\newblock {\em IEEE Transactions on Intelligent Transportation Systems\/}~{\bf 25\/}(2), 1913--1924.

\bibitem[\protect\citeauthoryear{Zhan}{Zhan}{2019}]{10.1785/0220190112}
Zhan, Z. (2019, 12).
\newblock {Distributed Acoustic Sensing Turns Fiber‐Optic Cables into Sensitive Seismic Antennas}.
\newblock {\em Seismological Research Letters\/}~{\bf 91\/}(1), 1--15.

\end{thebibliography}

\section{Mailing address}
\par
Jingxiao Liu, Haipeng Li, Siyuan Yuan, Biondo Biondi: 397 Panama Mall Mitchell Building, 3rd Floor, Stanford, CA 94305, United States.
\par
Jingxiao Liu, Hae Young Noh: Y2E2, 473 Via Ortega, Room 311,
Stanford, CA 94305, United States.

\section{Figure legends}
\par
{\bf Figure 1.} The DAS array section used in this study from the Stanford-2 DAS experiment and the DAS recording after preprocessing.
(a) Map view of a roadside section of the Stanford DAS-2 Array. 
Distances along the fiber are labeled on the map. 
Menlo Park is located to the west of the DAS array, and Stanford University is located to the east. 
The Spectral Analysis of Surface Waves (SASW) survey, performed in 2009, is approximately 200 meters east of our DAS array.
(b) A five-minute section of the DAS recordings. 
The dashed white lines indicate vehicle trajectories estimated using our Kalman filter and smoothing algorithm. 
The red boxes highlight the selected surface wave windows with isolated vehicles.
\par
{\bf Figure 2.} DAS signals induced by vehicles of different weights. 
(a) The distribution of the peak values of the averaged quasi-static signals (on a logarithmic scale). 
The peak values are used to represent the weights of vehicles.
Thresholds at the mode (0.57) and at the tail (1.2) of peak values differentiate between light, mid-weight, and heavy vehicles. 
(b) The quasi-static DAS signals induced by classified vehicles of different weights. 
The solid curves represent the mean signals, and the shaded areas indicate the signal ranges. 
(c) The power spectral density (PSD) of surface wave signals in the frequency band of 1--25 Hz for vehicles of different weights. 
The solid curves show the mean PSDs, and the shaded areas indicate the PSD ranges. 
(d)-(f) Examples of surface wave signals (1--30 Hz) for vehicles of different weights plotted with the same color map. 
The color version of this figure is available only in the electronic edition.
\par
{\bf Figure 3.} DAS signals induced by vehicles of different speeds. 
(a) The distribution of vehicle speeds estimated using the specialized Kalman filter and smoothing algorithm. 
Thresholds at the mean plus and minus the standard deviation of the vehicle speed differentiate between slow, mid-speed, and fast vehicles. 
(b) The quasi-static DAS signals induced by vehicles of different speeds. 
(c) The power spectral density (PSD) of surface wave signals for vehicles of different speeds. 
PSD values at most frequencies slightly increase as the vehicle speed increases. 
(d)-(f) Examples of surface wave signals for vehicles of different speeds.
The dashed lines represent the estimated vehicle motions.
The color version of this figure is available only in the electronic edition.
\par
{\bf Figure 4.} Retrieved VSGs and dispersion measurements from vehicles of different weights at the pivot location of 700 meters along the fiber (a grass lawn).
(a)–(c) The VSGs stacked from the same number (60 VSGs) of passing vehicles with light, middle, and heavy weights, respectively. 
(d)–(f) The computed dispersion spectra using the phase-shift method.
(g)–(i) The picked dispersion curves from (d)–(f). 
The blue lines show dispersion curves from VSGs constructed by stacking randomly sampled 60 vehicle-induced surface wave windows from each set of vehicles. 
The red dots represent the averaged shear wave velocities, with error bars indicating the standard deviation.
\par
{\bf Figure 5.} The same as Figure~\ref{fig:dispersion_size} but for vehicles of slow, middle, and fast traveling speeds.
\par
{\bf Figure 6.} Inversion results of near-surface structures at the 700-meter pivot location. 
(a)-(c) 1-D $V_S$ model of the near-surface structure for light, mid-weight, and heavy vehicles, respectively. 
Gray colors indicate the logarithmic misfits. 
Purple solid lines show the model parameter boundaries detailed in Table~\ref{tab:param}. 
Yellow and blue dashed lines represent the best model (detailed in Table~\ref{tab:best}) with the lowest misfit value and the mean of the top 30\% model. The green dashed line shows the velocity model from the SASW survey.
(d)-(f) Model fit of dispersion curves for light, mid-weight, and heavy vehicles, respectively. 
Blue dots with error bars are estimations from Figure~\ref{fig:dispersion_size}. 
Lines are model predictions of dispersion curves, including the top 70\% of the predicted models. 
The color version of this figure is available only in the electronic edition.
\par
{\bf Figure 7.} The same as Figure~\ref{fig:imaging_weights} but for vehicles of slow, middle, and fast traveling speeds. 
The color version of this figure is available only in the electronic edition.
\par
{\bf Figure 8.} Uncertainty quantification of the picked dispersion curves for different numbers of vehicles with varying weights and speeds. 
The uncertainties are quantified by bootstrapping the standard deviations of the estimated fundamental mode dispersion curves. (a) shows the standard deviations of the fundamental mode dispersion curves for vehicles with different weights at pivot locations of 700 meters.
(b) shows the standard deviations of the fundamental mode dispersion curves based on the percentage of vehicles' VSGs stacked with varying vehicle weights.
(c) displays the converged standard deviations of the fundamental mode dispersion curves at various frequencies for vehicles with different weights.
(d), (e) and (f) present the same data as (a), (b), and (c) but for vehicles with different speeds. 
The color version of this figure is available only in the electronic edition.
\par
{\bf Figure 9.}
Rayleigh surface‐wave phase velocity sensitivity kernels. 
The color version of this figure is available only in the electronic edition.

\end{document}